\documentclass[aps,prb,amsmath,amssymb,footinbib,showpacs,twocolumn,superscriptaddress]{revtex4-1}
\usepackage{amsmath}
\usepackage{amssymb}
\usepackage{amsthm}
\usepackage{setspace}
\usepackage{graphicx}
\usepackage{braket}
\usepackage{mathrsfs}
\usepackage{physics}
\usepackage{float}
\usepackage[colorlinks = true,linkcolor = red,urlcolor  = blue,citecolor = blue,anchorcolor = blue]{hyperref}
\usepackage[utf8]{inputenc}
\usepackage[english]{babel}
\usepackage{bm}
\usepackage{xcolor}
\usepackage{mathtools}

\begin{document}

\title{A Two-Kind-Boson Mixture Honeycomb Hamiltonian of Bloch Exciton-Polaritons} 

\author{Haining Pan}
\affiliation{Institute for Quantum Computing, University of Waterloo, 200 University Ave. West, Waterloo, ON, N2L 3G1, Canada}
\affiliation{Condensed Matter Theory Center and Joint Quantum Institute, Department of Physics, University of Maryland, College Park, MD 20742, USA}
\author{K. Winkler}
\affiliation{Technische Physik, Physikalisches Institut, Universit\"at W\"urzbug, D-97074 W\"urzburg, Germany}
\author{Mats Powlowski}
\affiliation{Institute for Quantum Computing, University of Waterloo, 200 University Ave. West, Waterloo, ON, N2L 3G1, Canada}
\affiliation{Department of Electrical and Computer Engineering, University of Waterloo, 200 University Ave. West, Waterloo, ON, N2L 3G1, Canada}
\author{Ming Xie}
\affiliation{Department of Physics, The University of Texas at Austin, Austin, TX 78712, USA}
\author{A. Schade}
\author{M. Emmerling}
\author{M. Kamp}
\author{S. Klembt}
\author{C. Schneider}
\affiliation{Technische Physik, Physikalisches Institut, Universit\"at W\"urzbug, D-97074 W\"urzburg, Germany}
\author{Tim Byrnes}
\affiliation{State Key Laboratory of Precision Spectroscopy, School of Physical and Material Sciences, East China Normal University, Shanghai 200062, China}
\affiliation{New York University Shanghai, 1555 Century Ave., Pudong, Shanghai 200122, China}
\affiliation{NYU-ECNU Institute of Physics at NYU Shanghai, 366 Zhongshan Road North, Shanghai 200062, China}
\affiliation{Department of Physics, New York University, New York, NY 10003, USA}
\author{S. H\"ofling}
\affiliation{Technische Physik, Physikalisches Institut, Universit\"at W\"urzbug, D-97074 W\"urzburg, Germany}
\affiliation{School of Physics and Astronomy, University of St. Andrews, St. Andrews, KY16 9SS, United Kingdom}
\author{Na Young Kim}
\email{nayoung.kim@uwaterloo.ca}
\affiliation{Institute for Quantum Computing, University of Waterloo, 200 University Ave. West, Waterloo, ON, N2L 3G1, Canada}
\affiliation{Department of Electrical and Computer Engineering, University of Waterloo, 200 University Ave. West, Waterloo, ON, N2L 3G1, Canada}
\affiliation{Department of Physics and Astronomy, University of Waterloo, 200 University Ave. West, Waterloo, ON, N2L 3G1, Canada}\affiliation{Department of Chemistry, University of Waterloo, 200 University Ave. West, Waterloo, ON, N2L 3G1, Canada}

\date{\today}

\begin{abstract}
	
	The electronic bandstructure of a solid is a collection of allowed bands separated by forbidden bands, revealing the geometric symmetry of the crystal structures. Comprehensive knowledge of the bandstructure with band parameters explains intrinsic physical, chemical and mechanical properties of the solid. Here we report the artificial polaritonic bandstructures of two-dimensional honeycomb lattices for microcavity exciton-polaritons using GaAs semiconductors in the wide-range detuning values, from cavity-photon-like (red-detuned) to exciton-like (blue-detuned) regimes. In order to understand the experimental bandstructures and their band parameters, such as gap energies, bandwidths, hopping integrals and density of states, we originally establish a polariton band theory within an augmented plane wave method with two-kind-bosons, cavity photons trapped at the lattice sites and freely moving excitons. In particular, this two-kind-band theory is absolutely essential to elucidate the exciton effect in the bandstructures of blue-detuned exciton-polaritons, where the flattened exciton-like dispersion appears at larger in-plane momentum values captured in our experimental access window. We reach an excellent agreement between theory and experiments in all detuning values.	
\end{abstract}


\maketitle

\section{Introduction}\label{sec:intro}

When identical particles are brought in a perfectly periodic lattice potential, their degenerate energy states are reorganized into allowed energy bands separated by forbidden energy gaps due to the spatial orbital wavefunction overlap.~\cite{ashcroft1976solid}  An orbital overlap between neighboring particles is quantified by a hopping integral, which determines the bandwidth of the allowed energy bands, and the spectral energy gap appears proportional to the potential strength. These particles are beautifully represented by Bloch waves, solutions to a Hamiltonian with an effective single-particle periodic-potential. In order to resemble electrons in natural crystals, artificial periodic lattices are designed and created to engineer the strengths of the particles' orbital overlap and their interaction governed by lattice geometries. In particular, a honeycomb crystal structure appears ubiquitously in solids such as graphite,~\cite{wallace1947band} graphene,~\cite{castroneto2009electronic} carbon nanotubes,~\cite{saito1998physical} two-dimensional (2D) transition metal dichalcogenide,~\cite{geim2013van} and theoretical lattice models.~\cite{nussinov2015compass} Recently, tunable honeycomb lattices are synthesized to investigate massless Dirac energy dispersion and topological phases.~\cite{polini2013artificial} Exciton-polariton honeycomb lattices have been also created by etching methods, where bandstructures are measured and edge states are identified.~\cite{jacqmin2014direct, milicevic2017orbital,cerda-mendez2010polariton,milicevic2015edge,klembt2018ti} However, their work is limited to polaritons in a far-red-detuned regime, where an approximated photon description is sufficient. In this study, we realize the full bandstructures of Bloch exciton-polaritons in an artificial 2D honeycomb lattice at not only red-detuning ($\Delta < 0$) but also zero- and blue-detuning ($\Delta > 0$) values, resulting in different ratios of exciton and photon contributions. $\Delta$ is defined as the energy detuning between cavity photon and quantum-well exciton. This quantity controls the potential strength of Bloch exciton-polaritons, consequently, the band parameters such as bandgap energies, bandwidths, energy density of states and their hopping integrals.  

Microcavity exciton-polaritons are dual-quasiparticles of photon-dressed excitons as a manifestation of strong light-matter coupling in a monolithic quantum well-microcavity structure.~\cite{weisbuch1992observation,carusotto2013quantum}  As composite bosons in a low-density limit, obeying Bose-Einstein statistics, exciton-polaritons exhibit macroscopic coherence above quantum degeneracy threshold via stimulated scattering process originating from exchange interactions.~\cite{deng2002condensation,kasprzak2006bose,balili2007boseeinstein,byrnes2014exciton} There are several methods to produce a lateral confinement for trapping exciton-polaritons by modulating spatially either photons or excitons.  A thin-metal film deposition,~\cite{lai2007coherent} an etching for a pillar,~\cite{bajoni2008polariton} and a partial-etching and overgrowth~\cite{eldaif2006polariton} are implemented to pattern photonic lattices. These methods extend to produce 2D exciton-polariton lattices~\cite{angelakis2017quantum} with limited tuning parameters to observe high-orbital condensation,~\cite{kim2011dynamical, kim2013exciton, kim2014band} spin-orbit coupling in the Lieb Lattice,~\cite{whittaker2018exciton} Dirac cones,~\cite{jacqmin2014direct} and condensation in an energy gap~\cite{winkler2015polariton} and to explore magnetic orders.~\cite{ohadi2017spin,berloff2017realizing}  Theoretically, the exciton-polariton Hamiltonian in the artificial lattices can be mapped to that of the Hubbard model, with which Mott transition can be explored in exciton-polariton lattices.~\cite{byrnes2010mott}  In order to construct such Hubbard Hamiltonian, a first step is to engineer its energy terms: an on-site interaction energy and a kinetic energy associated with a hopping integral.~\cite{mahan2000manyparticle} Quantifying these terms is directly linked to the bandstructures and their parameters. Here we alter the design parameters to vary the lattice potential strength, from which we quantify values of band parameters experimentally. We achieve the complete understanding of the engineered polaritonic bandstructures by developing our two-kind-boson mixture polariton band theory. The two-kind-boson band theory explicitly describes free quantum-well excitons and bound cavity photons in artificial honeycomb periodic potentials as well as the strong dipole interaction between excitons and photons. The calculated bandstructures from our two-kind-boson band theory are in an excellent agreement with the experimental ones at all detuning values. Especially, the two-kind-boson band theory is correct to explain the bandstructurues of the blue detuned devices.

The paper is organized as follows: Section~\ref{sec:exp} describes our device and experimental setup and presents measured bandstructures with varying experimental parameters. In Sec.~\ref{sec:thy}, we introduce a complete two-kind-mixture Hamiltonian to compute theoretical bandstructures by taking into account experimental parameters. We further discuss the comparison of theory and experiments and quantify band parameters in Sec.~\ref{sec:discussion}.

\section{Experiment}\label{sec:exp}
		
\begin{figure}[htbp]
\centering  
\includegraphics[width=\columnwidth]{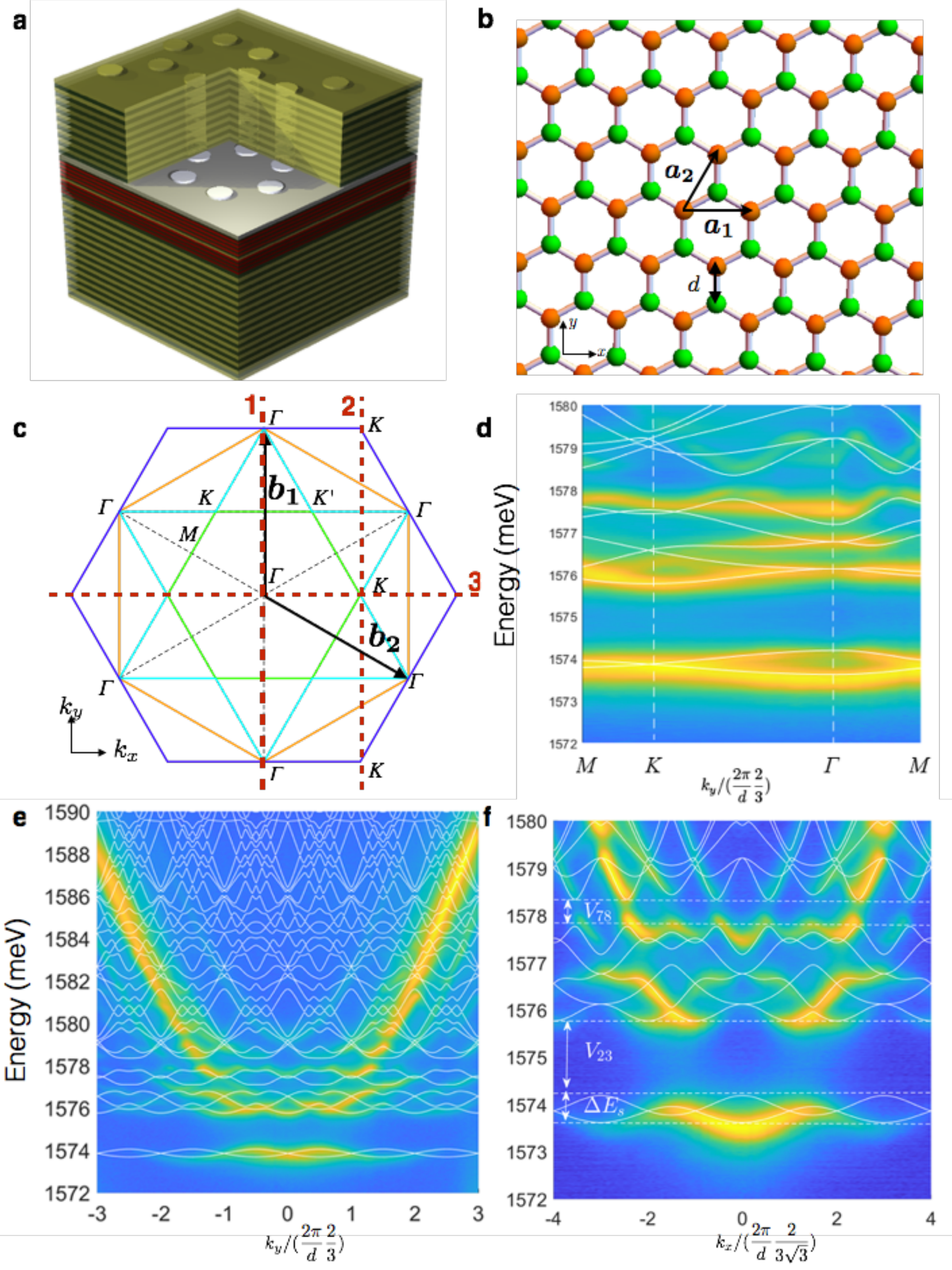}
\caption{ 
(a) An illustration of prepared honeycomb lattices patterned by the etching-overgrowth method. Each site is addressed as a circular disk, whose lithographic diameter is 2 $\mu$m and etching height is approximately 5 nm.
(b) A real-space honeycomb lattice is sketched by two unit vectors, $\bm{a}_1$ and $\bm{a}_2$, where two sublattices are colored green and orange.  The site-to-site distance is denoted as $d$. 
(c) The first four Brillouin zones (BZs) in an associated reciprocal space are built upon the reciprocal unit vectors  $\bm{b}_1 = (0, \frac{4\pi}{3d})$ and  $\bm{b}_2 = (\frac{2\pi}{\sqrt{3}d}, -\frac{2\pi}{3d})$ with high symmetry points $\it{\Gamma}$, $K$, $K^\prime$and $M$. 
(d) A folded representation of multiple BZs taken from experimental data ($\Delta = -$18.2 meV) along the high-symmetry points  denoted in (c). The theoretical folded BZs are overlaid on top of the experimental data in white straight lines. Note that the momentums are projected to the $ k_y $ direction.
(e) A representative measured bandstructure of the $d$ = 3 $ \mu $m device at  $\Delta= -18.2$  meV along  higher BZ regions, the vertical $ K-K$ line labeled as line 2 in (c), where the Dirac cones are seen in the lowest energy bands. 
(f) Another cross-sectional band structure is drawn along the line 3 in (c). Two distinct forbidden energy gaps are defined: $V_\text{23}$ and $V_\text{78}$, the energy gap between the second and third bands and the seventh and eighth bands, respectively.   The bandwidth of the lowest band $\Delta E_\text{S}$ is specified.  The computed energy bandstructures are shown in the straight white line in a repeated zone scheme, and yellow and blue in the color scale bar indicate high and low intensity data, respectively,  in (d), (e) and (f).
}
\label{fig:1}
\end{figure}

\begin{figure*}
 \centering 
 \includegraphics[width=\textwidth]{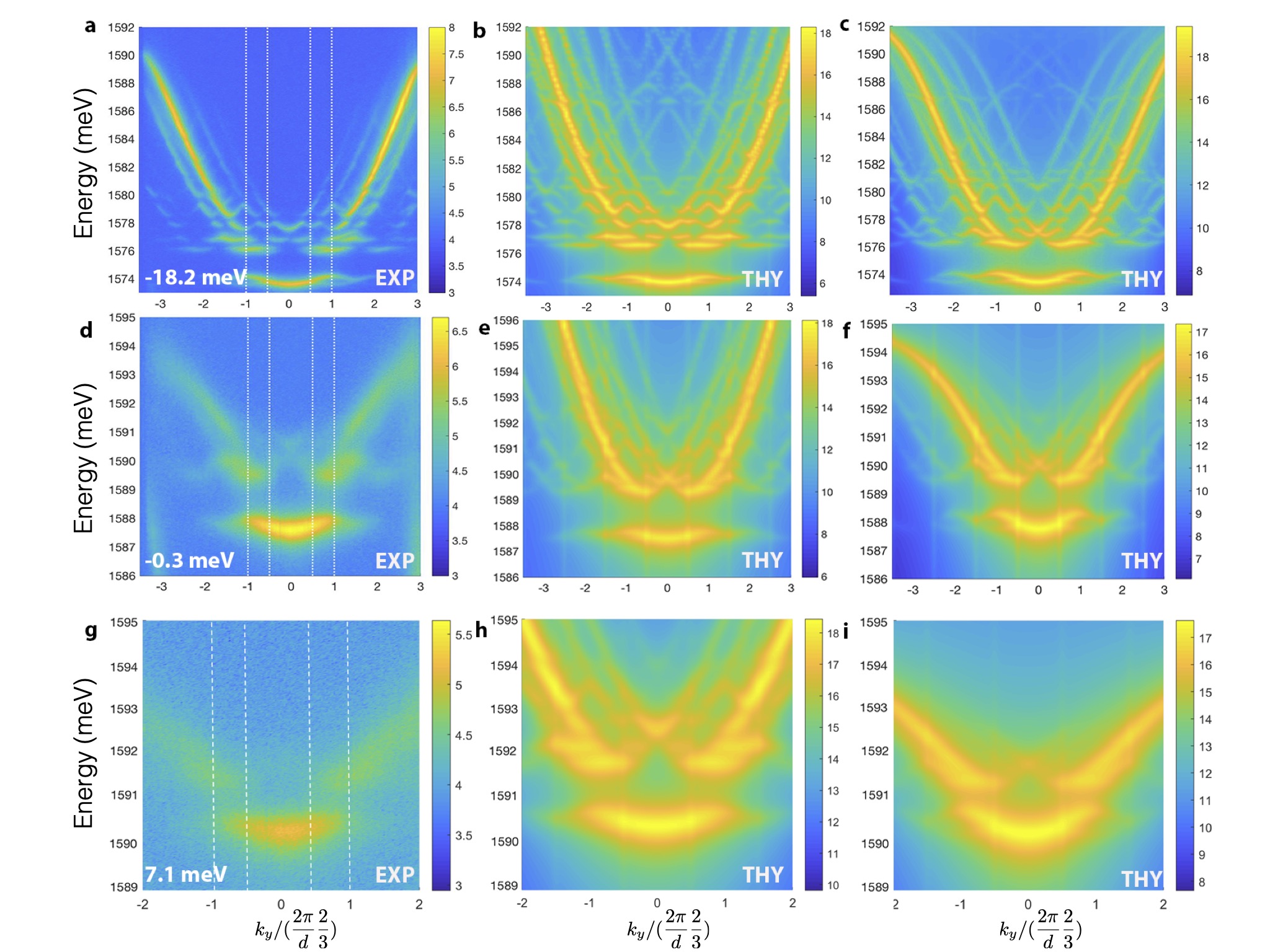}
 \caption{Direct comparison of  experimental bandstructures (left column, (a), (d), (g)) with calculated bandstructures by two different methods, the approximated $\tilde{\bm{H}}_1$ (middle column, (b), (e), (h)) and the complete Hamiltonian $\bm{H}$ (right column,(c), (f), (i)) at three different detuning values:  a red-detuned device with  $\Delta = -18.2$  meV (a),  a zero-detuned device with $\Delta = -0.3$ meV (d), and a blue-detuned device with  $\Delta = 7.1$  meV (g). These data are taken along the line 1 marked in Fig. 1(c). The vertical dotted lines in the experimental bandstructures indicate the zone boundaries.}
\label{fig:2}
\end{figure*}
In this study, we used a wafer which is composed of two stacks of four 7 nm-thick GaAs quantum-wells embedded in a $\lambda$/2-AlAs cavity structure sandwiched by 32-pair top and 37-pair bottom distributed Bragg reflectors, alternating AlAs and Al$_{0.2}$Ga$_{0.8}$As layers. The spatial cavity length variations over the wafer vary detuning values $\Delta$ from $ -18.2 $ meV to $ 7.1 $ meV in our sample. Detuning values are computed from $\Delta = E_c (k_\parallel = 0) - E_X (k_\parallel = 0) $, where $E_c$ and $E_X$ are independent photon and exciton energy.  An individual block has a pattern of a honeycomb array of circular disks by etching the top cavity layer, whose topology is consecutively translated to subsequent upper layers during overgrowth.~\cite{eldaif2006polariton, winkler2015polariton}.  

Each circular disk of the honycomb lattice has a fixed diameter to be 2 $\mu$m, and its etching depth is set to be 5 nm. The cavity-layer thickness variations modulate the photonic energies in space to create a periodic potential whose amplitude ranges 1-5 meV. The resulting sample schematic is sketched in Fig.~\ref{fig:1}(a). The site-to-site distance $d$ of the neighboring sites has values of 3 and 4 $\mu$m, and each block for a specific $d$ has the size of a 150 $\mu$m-by-150 $\mu$m.  A unit-cell in the real-space honeycomb lattice potential is defined by primitive unit vectors $\bm{a}_1$  and $\bm{a}_2$, where the nearest-neighbor distance is $d$ denoted in Fig.~\ref{fig:1}(b). Their reciprocal lattice vectors $\bm{b}_1$  and  $\bm{b}_2$  construct 2D hexagonal Brillouin zones (BZ) with rotational symmetry points, $\it{\Gamma}, K, K^\prime$, and $M$ (Fig.~\ref{fig:1}(c)). We make 8 blocks with $d$ = 3 $\mu$m in the range of detuning $\Delta \in$ ($ -18.2 $ , 7.1) meV, whose corresponding photonic fraction values, $|C(k_\parallel = 0)|^2$ lie between 0.92 and 0.24, where $C$ is the Hopfield coefficient and $k_\parallel$ is the in-plane momentum (We will also use $ k $ to denote $ k_\parallel $ if no potential confusion will occur in the following text). 

The sample containing many blocks is cooled down to 4 - 6 K, and is excited by a continuous-wave laser at the fixed wavelength 1.616 eV (767.205 nm) at the angle of 60-degrees. Our detection is not polarization-selective. We keep the laser power to be 0.1 - 1 mW, which is much lower than the threshold pump power values ~ 40 - 60 mW at various detuning positions. The laser spot is oval-shaped due to the finite-angle pumping scheme with a size of about 120 $\mu$m-by-60 $\mu$m. 

The standard angle-resolved photoluminescence spectroscopy allows us to map the bandstructures of the honeycomb lattice in the extended zone scheme, and its folded-zone representation is presented in Fig.~\ref{fig:1}(d) along the three high-symmetry points ($\it{\Gamma}, K, M$) drawn in Fig.~\ref{fig:1}(c).
We plot representative experimental bandstructures taken at two distinct lines, line 2 (Fig.~\ref{fig:1}(e)) and line 3 (Fig.~\ref{fig:1}(f)) defined in Fig.~\ref{fig:1}(c). The full massless Dirac dispersions of the $s$-bands are captured in the second BZ cutting through $K$ or $K^\prime$ in Fig.~\ref{fig:1}(e), while only the lowest Dirac band is seen in the first BZ along $\it{\Gamma} - K$ (or $K^\prime$) (Fig.~\ref{fig:1}(f)).  
	
Figure~\ref{fig:2} presents representative experimental polaritonic bandstructures ($d$ = 3 $\mu$m) taken along the line 1 in Fig.~ \ref{fig:1}(c) at three different detuning values $\Delta = -18.2 $ meV (a), $-0.3$ meV (d), and 7.1 meV (g), where the strong $\Delta$-dependence is observed. As $\Delta$ becomes positive, higher exciton fraction reduces the energy window from $\sim$ 17 meV (Fig.~ \ref{fig:2}(a)) to $\sim$ 5 meV  (Fig.~ \ref{fig:2}(g)). Consequently, the number of accessible bands within the energy window decreases as $\Delta$ is more positive. Noticeably, the flattened bands at the large momentum values are more visible within our detected angular regions. All of these behaviors are associated with the heavier effective mass of the resulting exciton-polaritons arising from the higher fraction of excitons as $\Delta$ increases. In addition, the spectral linewidths of individual bands become much broader in the blue-detuned devices (Fig.~ \ref{fig:2}(g)), resulting in the smaller forbidden energy band regions, which blur sharp separation of different bands. 

\section{Theoretical model}\label{sec:thy}

In order to explain experimental exciton-polariton bandstructures, we develop theoretical models by solving a Hamiltonian within an augmented plane-wave method.  We first apply an approximated Hamiltonian with the effective mass of exciton-polaritons, which has been working well for photon-like exciton-polaritons in previous works.~\cite{kim2011dynamical, kim2013exciton, jacqmin2014direct}  The theoretically computed bandstructures obtained from the approximated Hamiltonian are displayed in Figs.~\ref{fig:2}(b)(e)(h). When we calculate the best fitted theoretical plots matching to experimental data, we have only three fitting parameters: the effective polariton potential strength  $V_{\text{eff}}$ as a function of $\Delta$, effective mass of exciton-polaritons $ m^* $, and site-to-site distance $d$. Our theoretical bandstructures plots also take into account experimental linewidths in energy and momentum. The intensity of each state at a particular momentum $\bm{k}$ is computed by projecting the Bloch wavefunction onto a free-moving photon and integrating over the first BZ. We also take into account the spectral and momentum linewidth information from experimental bandstructures. 

Most features of allowed energy bands and gaps within the first BZ are reproduced very well in all three detuning regimes. In the red-detuned device, experimental data (Fig. \ref{fig:2}(a)) and theoretical plots by this method (Fig. \ref{fig:2}(b)) are in excellent agreement within the region of $|k_y| \leq 3 (\frac{2\pi}{d}\frac{2}{3})$. A slight difference between two is visible in a region of very large $k_y$ values ($> 3 (\frac{2\pi}{d}\frac{2}{3})$) within our optical access window determined by the numerical aperture (NA = 0.55) of our objective lens. However, this approximated Hamiltonian fails to explain the bandstructures in two other devices, where the  discrepancy between experiments and theory becomes dramatically noticeable in devices (Figs.~\ref{fig:2}(d) and (g), $\Delta = -0.3$ and $ 7.1 $ meV, respectively). The flattened nature of the lower-polariton (LP) higher bands at larger $k_y$ values originates from the bare exciton dispersion. The composition of exciton and photon fractions in the strongly-coupled LP dispersions depends on not only the $\Delta$ values, which is here defined at  $k_\parallel$ = 0 but also the non-zero $k_\parallel$-values. Near $k_\parallel$= 0, the photonic component is stronger, while the excitonic component is stronger near large $k_\parallel$-values. Therefore, the LP dispersion is photon-like near smaller  $k_\parallel$-values, while the LP dispersion is exciton-like at larger  $k_\parallel$ values. The cross-over  $k_\parallel$ value between two exciton-like and photon-like regimes  within the LP dispersion is determined by $\Delta$. For the blue detuned device, this cross-over $k_\parallel$-value is smaller so that it appears within our observation range determined by the objective numerical aperture, whereas this cross-over $k_\parallel$-value is near the edge of or beyond our observation access for the red detuned devices so that the dispersion looks parabolic.  

For differently behaving two-kind entities, the confined cavity photons trapped by the engineered potential and the free excitons, we originally introduce the complete Hamiltonian, $\bm{H}$, which is divided into three terms: the exciton Hamiltonian $\bm{H}_X$, the cavity-photon Hamiltonian, $\bm{H}_c$, and the exciton-photon coupling Hamiltonian, $\bm{H}_{X-c}$, i.e. 
\begin{equation}\label{key}
\bm{H} = \bm{H}_X + \bm{H}_c + \bm{H}_{X-c}.
\end{equation}
In momentum space, $\bm{H}_c$ is expressed as
\begin{widetext}
\begin{equation}
	\bm{H}_c =\left(\varepsilon_c + \frac{\hbar^2 (\bm{k}+\bm{G}_{h, h^\prime})^2}{2m_c}+V(\bm{G}_{h, h^\prime, h'', h'''})\right)  \bm{c}^\dagger_{\bm{k}+\bm{G}_{h, h^\prime}}\bm{c}_{\bm{k}+\bm{G}_{h'', h'''}},
\end{equation}
\end{widetext}
where $\varepsilon_c$ is the cavity-photon energy offset, the effective mass of cavity-photon is $m_c$,  and $\bm{c}^\dagger_{\bm{k}}$ and $\bm{c}_{\bm{k}}$ are the cavity-photon creation and annihilation operators at momentum $\bm{k}$. 
$\bm{G}_{h, h^\prime}$ is the general reciprocal lattice vectors, expanded by $\bm{b}_1$  and  $\bm{b}_2$, such that $\bm{G}_{h, h^\prime} = h \bm{b}_1 + h^\prime \bm{b}_2$. The coefficients of $h,h'$ are the indices of 2D Fourier transformation. Note that  $V(\bm{G}_{h, h^\prime})$ is the Fourier transformation of the honeycomb lattice in real space, formed by the honeycomb arrays of $V(\bm{r})$ at each site,
\begin{equation}\label{key}
V(\bm{r})=-V_0 \cdot \theta (|\bm{R}-\bm{r}|),
\end{equation}
where $V_0$ is the maximum potential strength for the 100$\%$ cavity-photon, approximately 5 meV, and $\bm{R}$ is the radius of the circular potential well, $ \theta(x) $ is Heaviside step function.  On the other hand, $\bm{H}_X$ has only the kinetic energy term, reading
\begin{equation}\label{key}
\bm{H}_X = \Big( \varepsilon_X + \frac{\hbar^2 (\bm{k}+\bm{G}_{h, h^\prime})^2}{2m_X} \Big) \bm{a}^\dagger_{\bm{k}+\bm{G}_{h, h^\prime}}\bm{a}_{\bm{k}+\bm{G}_{h, h^\prime}},
\end{equation}
where  $\varepsilon_X$ is the exciton energy offset, $m_X$ is the effective mass of an exciton, and  $\bm{a}^\dagger_{\bm{k}}$ and  $\bm{a}_{\bm{k}}$ are the exciton creation and annihilation operators at momentum $\bm{k}$. $\bm{H}_{X-c}$ indicates the energy interchange of exciton and cavity photon, 
\begin{equation}\label{key}
\bm{H}_{X-c} = g_0  (\bm{a}^\dagger_{\bm{k}+\bm{G}_{h, h^\prime}} \bm{c}_{\bm{k}+\bm{G}_{h, h^\prime}} +\bm{c}^\dagger_{\bm{k}+\bm{G}_{h, h^\prime}} \bm{a}_{\bm{k}+\bm{G}_{h, h^\prime}}),
\end{equation}
where $g_0$ = 5.75 meV is the strength of exciton-photon coupling. 

The theoretical bandstructures by the complete $\bm{H}$ are presented in Figs.~\ref{fig:2}(c)(h)(i). In comparison to the experimental bandstructures as well as the approximated Hamiltonian calculation, we conclude that, only for extremely photon-like exciton-polaritons, the preliminary plane wave basis method can be still valid since the photon fractions are higher at almost all accessible momentum values. However, for more exciton-like exciton-polaritons, the complete Hamiltonian is required because the effect of free excitons are no longer negligible. We can also show the mathematical details of the relation between the approximated and the complete Hamiltonians in Appendix \ref{app:A}.

\begin{figure}[htbp]
	\centering  \includegraphics[width=\columnwidth]{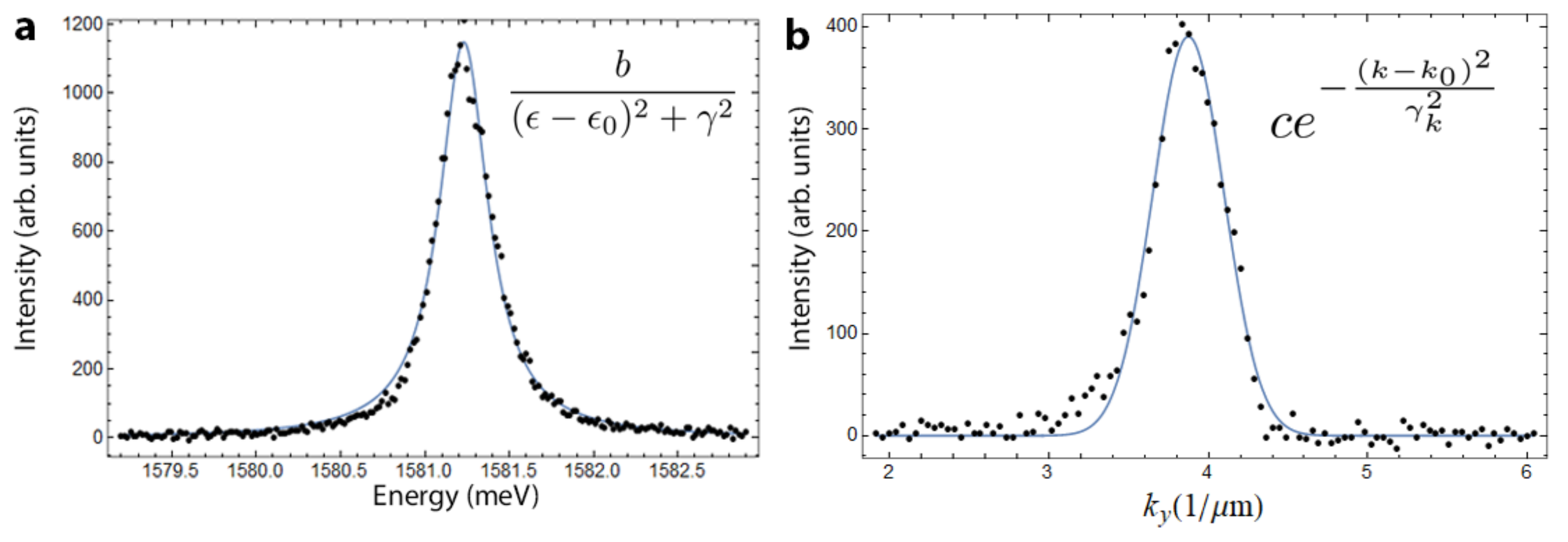}
	\caption{(a) The Lorentzian model fits our experimental data in the energy axis. The experimental data are taken at $\it{\Gamma}$. The fitting parameters of this particular plot read $(\epsilon_0, \gamma, b) \rightarrow (1581.23, 0.18, 35.44)$ in unit of (meV, meV, meV$^2$). (b) The Gaussian model fits our experimental data in the momentum axis. The experimental data are taken at where $ E = 1591.3$ meV. The fitting parameters of this particular plot read $(c, k_0, \gamma_k) \rightarrow (390.51, 4.02, 0.15)$, where $k_0$ and $\gamma_k$  have a unit of 1/$\mu$m.
	}
	\label{fig:3}
\end{figure}

Since the signals in photoluminescence experiments in Fig.~\ref{fig:2} are the intensity of LPs, we consider the population of LPs in theoretical plots. We assume that LPs trapped in the periodic lattice potential have been entirely converted to photons in the cavity. Therefore, we define the expected leakage LP emission intensity to be proportional to the projection from the Bloch wave function $ \psi_{n,\bm{k_\parallel}} $ to a free-moving photon $ \phi_{\bm{k_\parallel}}=e^{i\bm{k_\parallel} \cdot \bm{r}} $, namely, intensity \cite{lai2007coherent}
\begin{equation}\label{eq:rho}
\rho(n,\bm{k_\parallel})\propto\abs{\innerproduct{\phi_{\bm{k_\parallel}}}{\psi_{n,\bm{k_\parallel}}}}^2,
\end{equation}
where $ \bm{k_\parallel} $ is the in-plane lattice momentum.

The theoretical bandstructures presented in Fig.~\ref{fig:2} of the main text contain the spectral and momentum linewidth information from experimental bandstructures. Figure~\ref{fig:3} shows representative data of linewidth extractions in energy and momentum from a $d$ = 3 $\mu$m device at $\Delta = {-9.6}$ meV. The energy plot fits well with the Lorentzian equation (Fig.~\ref{fig:3}(a)), whereas the momentum cross-sectional plot works well with the Gaussian-shape fit (Fig.~\ref{fig:3}(b)). Therefore the overall emission intensity in Eq.\eqref{eq:rho} is modulated by another Lorentizian term along energy and Gaussian term along momentum. Namely, the overall intensity of the plot in the whole range at the momentum $\bm{k}$ and the energy $E$  is given by

\begin{equation}
I (E, \bm{k}) = \sum_i \sum_{\bm{h}} \sum_{\bm{k}_0} |a^i_{\bm{h}} (\bm{k})|^2 \frac{\exp{-\left(\frac{\bm{k}-\left(\bm{k}_0+\bm{h}\cdot\bm{b}\right)}{\gamma_k}\right)^2}}{|E - E_i(\bm{k})|^2+\gamma^2},
\end{equation}

with the energy and momentum relaxation rates $\gamma, \gamma_k$, respectively for the $i$-th energy state. $a^i_{\bm{h}}$ indicates component of eigenstate which corresponds $G_{\bm{h}}$ in the $i$-th energy state, where $ \bm{h}=\left(h_1,h_2\right) $ is the Fourier expansion order, $ h_1 $ and $ h_2 $ varying from $(-n, -n+1,..., n-1, n)$. $ \bm{b} $ is the reciprocal unit vector, given as $ \bm{b}=\left(b_1,b_2\right)=\left(\left(0,\frac{4\pi}{3d}\right),\left(\frac{2\pi}{\sqrt{3}d},-\frac{2\pi}{3d}\right)\right) $. $\bm{k}_0$ are the reference points taken from the experimental energy-dispersion relations, which are uniformly distributed in the first BZ. We typically take 25 points in each BZ in order to prevent the coarse granularities. Note that in Fig.~\ref{fig:2}, we plot the intensity actually in log scale, $\log(I(E,\bm{k}))$ in order to show the contrast between high and low intensity clearly. 

\begin{figure}[htbp]
 \centering \includegraphics[width=9cm]{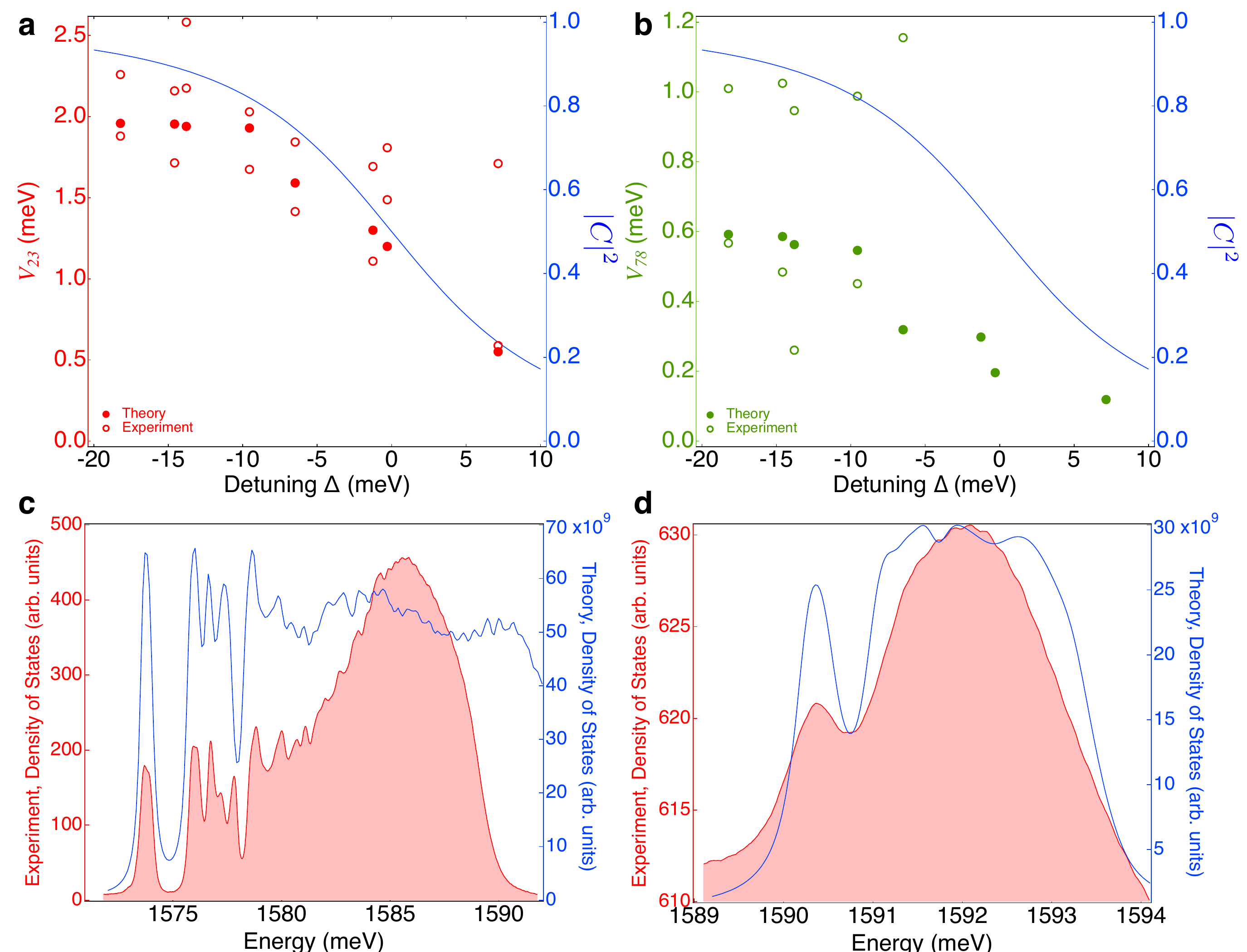}
\caption{$\Delta$-dependent gap energies $V_{\text{23}}$ (a) and $V_{\text{78}}$ (b) between the second and the third states and between the seventh and the eighth states, which are defined in Fig.~\ref{fig:1}(f). The photonic fraction values, $|C|^2$, are drawn against different $\Delta$ values. The filled circles present theoretical values from the Hamiltonian calculations, whereas experimentally, the gap energy values are taken from two methods shown in open circles. The lower bounds are taken from $ \text{min}(E_{i+1})- \text{max}(E_{i}) $, whereas the upper bounds are from $ \text{min}(E_{i+1}-E_{i}) $, where $ \text{min}$ and $\text{max}$ are functions to take the minimum and maximum values. Note that we cannot extract the lower and upper bound of gap energies from experiments in blue-detuned devices due to the finite resolution and small energy windows. Normalized energy density of states are drawn from experimental (red) and theoretical (blue) band structures at $\Delta = -18.2 $ meV (c) and $\Delta =  7.1 $ meV (d).
 }
\label{fig:4}
\end{figure}

\section{Discussion}\label{sec:discussion}
Now we examine $\Delta$-dependent bandstructure parameters: the gap energy values and energy density of states (DOS) in Fig.~\ref{fig:4}.  $V_{\text{23}}$ and $V_{\text{78}}$ are explicitly indicated in Fig.~\ref{fig:1}(f), which separate between the highest $s$-band and the lowest $p$-band at zone boundaries and between the lowest $d$-band and the second lowest $d$-band, respectively.  The reason we look at these two energy gaps is that the gaps exist at all momentum values, while $V_{\text{67}}$ between the lowest $p$-band and the lowest $d$-band disappears at certain momentum values. Theoretical gap energy values are extracted from the Hamiltonian solutions denoted as filled circles in Fig.~\ref{fig:4}(a) and (b). Experimentally, owing to the spectral linewidths of bands, we take these values with two methods: One values are determined by projecting only peak values at a given wavenumber to the energy axis and finding the gapped regions, which are in the lower side, while other values are obtained by projecting all intensities to the energy axis and finding the distance between the peaks of the allowed bands, which often are higher values. These two values are drawn in open circles as a function of $\Delta$ in Fig.~\ref{fig:4}(a) and (b). Since the gap energy values are proportional to the effective potential strength, at different device locations, the trapping potential strength linearly increases with $|C|^2$. In other words, the more photon-like exciton-polaritons encounter the stronger trapping potential, leading to wider gap openings at the zone boundaries. The $V_{\text{23}}-\Delta$ and $V_{\text{78}}-\Delta$ trends follow very well with the $|C|^2-\Delta$ relation, where $|C|^2$ is the photonic fraction at the global minimum,  $k_\parallel$ = 0.

We also plot the energy DOS by integrating the intensities over momentum values at two extreme $\Delta$ values in Figs.~\ref{fig:4}(c) and (d) with both experimental and theoretical bandstructures. In solid-state systems, the DOS is a useful quantity to explain particle motions and compute various physical parameters such as particle numbers combining with appropriate particle distribution functions. The DOS of the red-detuned device exhibits a series of peaks and the bounded lowest band is isolated by the gapped regions clearly visible in Fig.~\ref{fig:4}(c). However, the DOS of the blue-detuned device consists of two broad regions with occupied energy states.

\begin{figure}
 \centering  \includegraphics[width=9cm]{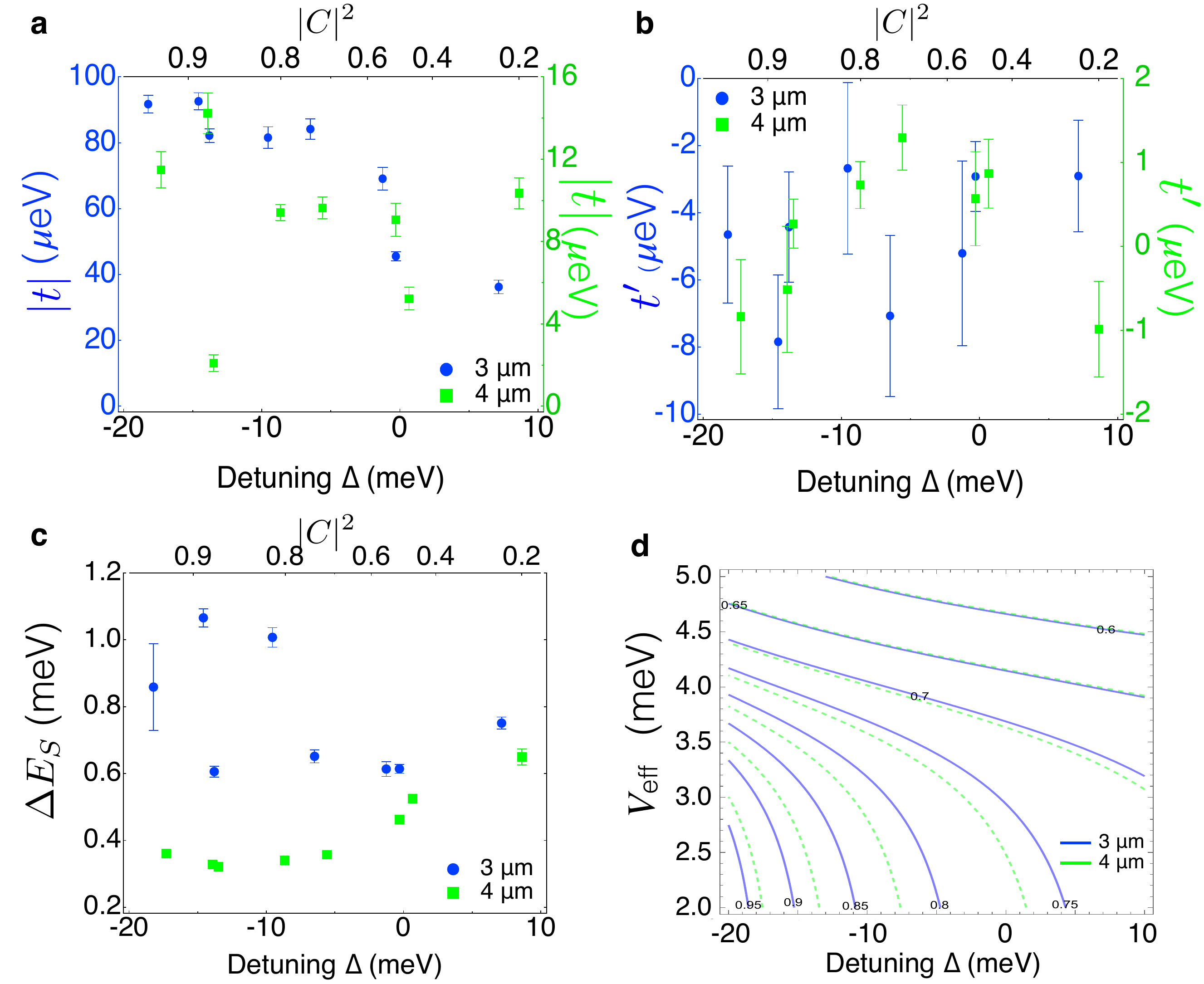}
\caption{ The tight-binding Hamiltonian fitting results of the nearest-neighbor hopping integral  (a) and the next-nearest-neighbor hopping integral  (b) from the $d$ = 3 and 4 $\mu$m devices directly. (c) The energy bandwidth of the lowest $s$-band $\Delta E_{\text{S}}$ defined in Fig.~\ref{fig:1}(f). The error bars are the standard deviation of the full-width at half-maximum (FWHM). (d) The $\Delta$ and $V_{\text{eff}}$ effect of delocalization. The contour plot of the spatial wavefunction FWHMs ($ \mu $m) is computed for the $d$ = 3 (blue straight line) and 4 (green dotted line) $\mu$m devices. The contour values start from 0.9 (0.95) in the upper right corner for  $d$ = 3 (4)  $\mu$m and end as 0.6 in the lower left corder for both  $d$ = 3 and 4 (green dotted line) $\mu$m.
 }
\label{fig:5}
\end{figure}

For the bounded isotropic $s$-bands, we apply the tight-binding approximation with two fitting parameters, the nearest-neighbor hopping integrals $t$ and the next-nearest-neighbor hopping integrals $t^\prime$ (see Appendix \ref{app:tb}). The energy dispersion is simply written in terms of the $f$-function  with the three neighbor sites, and the fitting results of  $t$ and $t^\prime$ from $d$ = 3 $\mu$m (blue circle) and $d$ = 4 $\mu$m (green square) are collected in Figs.~\ref{fig:5}(a) and (b). The $t$-values for the $d$ = 3 $\mu$m devices monotonically decrease in the positive $\Delta$ sides, but the  $t$-values for the $d$ = 4 $\mu$m devices are about 5-6 times smaller than those for the  $d$ = 3 $\mu$m devices, and the $\Delta$-dependence is weak. In addition, the amplitudes of $t^\prime$ are 10-20 times smaller than those of $t$, indicating the weak overlaps between the next-nearest-neighbor sites. 

The monotonic decrease of the $t - \Delta$ relation in the $d$ = 3 $\mu$m devices is qualitatively similar to the $\Delta E_{\text{S}}-\Delta$ plot of the $d$ = 3 $\mu$m devices in Fig.~\ref{fig:5}(c). The bandwidth of the lowest $s$-band, $\Delta E_{\text{S}}$, reflects the strength of the overlap integral between neighboring sites. The greater the overlap is, the stronger curvature emerges, thus the wider band. In the positive $\Delta$ values, since exciton-polaritons are much heavier, they tend to be likely localized, reducing the overlap integral. In order to quantify the delocalization degree of exciton-polaritons, we compute how the $s$-band wavefunctions spatially spread from the center of the trap in Fig.~\ref{fig:5}(d) as a function of $\Delta$ and $V_{\text{eff}}$. These 2D plots teach that the photon-like exciton-polaritons tend to delocalize in a weaker trapping potential, whereas the exciton-like exciton-polaritons are localized in a trap. The difference between $d$ = 3 $\mu$m and $d$ = 4 $\mu$m is bigger in the negative $\Delta$ and shallower $V_{\text{eff}}$. This is partly because the more distant the traps are, the more isolated. Thus, we conclude that, when $d$ = 3 $\mu$m, the $s$-bands reveal the strong overlap integral between the wavefunctions, while for $d$ = 4 $\mu$m, $\Delta E_{\text{S}}$ is limited by exciton-polariton lifetimes.  

\section{Conclusion}\label{sec:conclusion}
We successfully engineer artificial bandstructures of exciton-polaritons in a variability of exciton and photon fractions at different detuning values. The experimental bandstructures are completely understood by the two-kind-boson Hamiltonian, which requires to explicitly address both confined photons and free excitons. Our two-kind exciton-polariton lattice system is suitable to investigate a physical system, where both itinerant and localized particles coexist. Furthermore, the higher-orbital bands are accessible, which enables us to systematically examine various orbital physics in the photon-like and exciton-like regimes when they are placed at zone boundaries or near the Dirac cones in the honeycomb lattice. In addition, we may incorporate density and polarization engineering to prepare for Bloch exciton-polaritons that reveals spin order in the non-linear regime at exotic lattice geometries for studying the interplay of topology, spin and interaction. 

\begin{acknowledgments}
N.Y.K acknowledges Y. Yamamoto, T. Oka, K. Sota, and A. Burkov for fruitful discussions. H.P., M.P. and N.Y.K are supported by Industry Canada and the Ontario Ministry of Research $\&$ Innovation through Early Researcher Awards. This research was undertaken thanks in part to funding from the Canada First Research Excellence Fund. H. P. thanks Z. Xie for the travel support to visit IQC. M.P. is a recipient of the Richard and Elizabeth Master Graduate Entrance Scholarship and NSERC Canada Graduate Scholarships-Master's Program. K. W., A. S., M. M., M. K., S. K., C. S., S. H. receive the support from the State of Bavaria.
\end{acknowledgments}

\appendix
\setcounter{secnumdepth}{3}
\setcounter{equation}{0}
\setcounter{figure}{0}
\renewcommand{\theequation}{A-\arabic{equation}}
\renewcommand{\thefigure}{A\arabic{figure}}
\renewcommand\figurename{Figure}
\renewcommand\tablename{Appendix Table}

\section{Approximation of complete Hamiltonian}\label{app:A}
\begin{figure*}[htbp]
	\centering  \includegraphics[width=\linewidth]{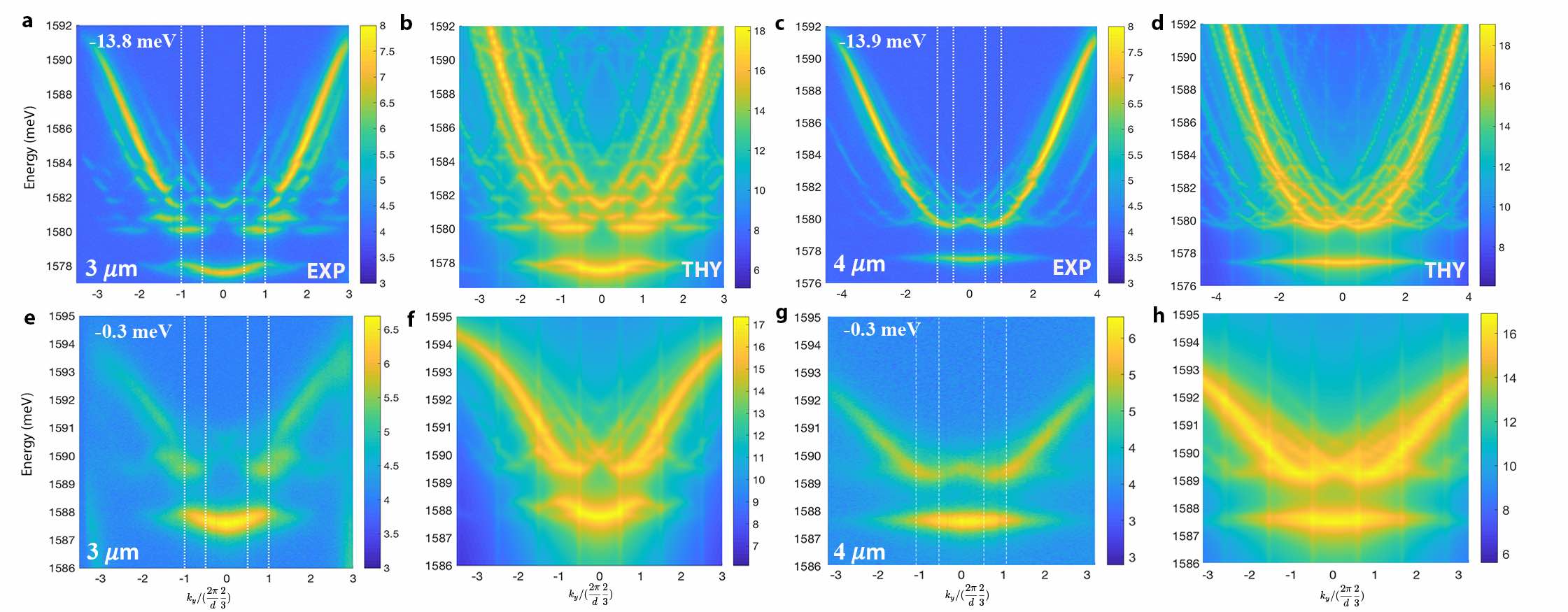}
	\caption{(a)(c)(e)(g), Cross-sectional energy dispersions along the dashed line 1 in Fig.~1(b) of the main text, passing through $\it{\Gamma} - M - \it{\Gamma}$.  Experimental data for $d$ = 3 and 4 $\mu$m  devices are presented in a log scale of the intensity in (a)(c). Corresponding theoretical bandstructures are computed and presented in a log scale in (b)(d). White vertical lines are drawn at $M = (0, \pm \frac{2\pi}{3d})$ in the first BZ and the $\it{\Gamma} =(0, \pm \frac{4\pi}{3d})$ of the second BZ.  Experimental (e)(g) and theoretical (f)(h) plots show detuning-value dependent bandstructures of honeycomb lattices.
		Theoretical bandstructures in (b) and (d) are computed by an approximated Hamiltonian, whereas those of (f) and (h) are from the complete Hamiltonian calculation with two individual components of cavity-photons and quantum-well excitons.
	}
	\label{fig:A1}
\end{figure*}
We have shown in the main text about the difference of complete Hamiltonian and approximated Hamiltonian. They differ much in the blue-detuned case, however, in the red-detuned limit ($\Delta \ll 0 $), the complete Hamiltonian presented above can be further approximated to a simpler single-component exciton-polariton Hamiltonian (Fig.~\ref{fig:A1}(a)-(h)). 

We first define an effective potential strength $V_{\text{eff}}$ , which exhibits the spatial variations at different  $\Delta$ values,
\begin{equation}
 V_{\text{eff}} = |C|^2 V_0.
\end{equation}
In the extreme red-detuned case, where we have $|C|^2 \sim 1$, the complete Hamiltonian can be approximated to the new Hamiltonian $\bm{H}_1$,
\begin{equation}
\bm{H}_1 = \frac{\hbar^2 (\bm{k}+\bm{G}_{h, h^\prime})^2}{2m_{\text{eff}}} + V_{\text{eff}} (\bm{G}_{h, h^\prime,h'', h'''}) + \varepsilon,
\end{equation}
where $\varepsilon$ is an energy offset value. The proof is as follows:
Note that the new Hamiltonian  $\bm{H}_1$ only includes the cavity-photon contribution with the effective mass $m_{\text{eff}}$, namely it has only the half-sized dimension of the complete Hamiltonian  $\bm{H}$. Therefore, we need to extend the space to the same dimension that
\begin{equation}
\tilde{\bm{H}}_1= \Big[\begin{array}{cc}
\bm{H}_1 & \bm{0} \\
\bm{0} & \bm{0} 
\end{array} \Big].
\end{equation}

Here, if we expand the Fourier series of the potential strength to $n$-th order $(-n, -n+1, \dots, n-1, n)$, we have $\bm{H}_1$ with the $(2n+1)\times(2n+1)$  dimension and  $\tilde{\bm{H}}_1$ with $2(2n+1)\times2(2n+1)$ dimensions. ($\bm{0}$ in above equation stands for the zero matrix). Now the complete Hamiltonian $\bm{H}$ and the new matrix $\tilde{\bm{H}}_1$ have the same dimension. The complete Hamiltonian  $\bm{H}$ explicitly reads
\begin{equation}
\bm{H}= \Big[\begin{array}{cc}
\bm{H}_c & \bm{g_0} \\
\bm{g_0} & \bm{H}_X 
\end{array} \Big],
\end{equation}
where
\begin{equation}
\bm{H}_c = \begin{bmatrix} 
\varepsilon_c + \frac{\hbar^2(\bm{k}+\bm{G}_{-n,-n})^2}{2m_c} & \dots & V(\bm{G}_{(-n,-n),(n,n)}) \\
\vdots & \ddots & \vdots \\
V(\bm{G}_{(n,n),(-n,-n)}) & \dots & \varepsilon_c + \frac{\hbar^2(\bm{k}+\bm{G}_{n,n})^2}{2m_c}\\
\end{bmatrix},
\end{equation}
\begin{equation}
\bm{H}_X = \begin{bmatrix} 
\varepsilon_X + \frac{\hbar^2(\bm{k}+\bm{G}_{-n,-n})^2}{2m_X} & \dots & 0 \\
\vdots & \ddots & \vdots \\
0 & \dots & \varepsilon_X + \frac{\hbar^2(\bm{k}+\bm{G}_{n,n})^2}{2m_X}\\
\end{bmatrix},
\end{equation}
and  $\bm{g_0}$ is the $(2n+1)\times(2n+1)$  diagonal matrix. Because of the mass of exciton $m_X \gg m_c$, the mass of cavity photon, the exciton kinetic term hardly contributes, consequently, this term just vanishes. By introducing the detuning energy $\Delta(\bm{k})$ , we have  $\Delta(k_\parallel = 0) = \varepsilon_c-\varepsilon_X$ (In the latter content, for the simplicity, we just use $\Delta$ to denote $\Delta(0)$). Finally, $\bm{H}_X$ is simplified to 
\begin{equation}
\bm{H}_X = \begin{bmatrix} 
\varepsilon_X  & \dots & 0 \\
\vdots & \ddots & \vdots \\
0 & \dots & \varepsilon_X \\
\end{bmatrix},
\end{equation}
and $\bm{H}_c$ reads
\begin{equation}
\bm{H}_c = \begin{bmatrix} 
\varepsilon_X + \Delta + \frac{\hbar^2(\bm{k}+\bm{G}_{-n,-n})^2}{2m_c} & \dots & V(\bm{G}_{(-n,-n),(n,n)}) \\
\vdots & \ddots & \vdots \\
V(\bm{G}_{(n,n),(-n,-n)}) & \dots & \varepsilon_X + \Delta+ \frac{\hbar^2(\bm{k}+\bm{G}_{n,n})^2}{2m_c}\\
\end{bmatrix},
\end{equation}
where  $\varepsilon_c$ is now replaced with $\varepsilon_X + \Delta$  for all diagonal elements by definition. If we take the component  $\varepsilon_X$ out of the diagonal elements of matrix,  $\bm{H}$ becomes
\begin{equation}
\bm{H} = \bm{\varepsilon}_X +  \begin{bmatrix} 
\tilde{\bm{H}}_c & \bm{g_0} \\
\bm{g_0} & \bm{0}\\
\end{bmatrix},
\end{equation}
where
\begin{equation}
\tilde{\bm{H}}_c = \begin{bmatrix} 
\Delta + \frac{\hbar^2(\bm{k}+\bm{G}_{-n,-n})^2}{2m_c} & \dots & V(\bm{G}_{(-n,-n),(n,n)}) \\
\vdots & \ddots & \vdots \\
V(\bm{G}_{(n,n),(-n,-n)}) & \dots & \Delta+ \frac{\hbar^2(\bm{k}+\bm{G}_{n,n})^2}{2m_c}\\
\end{bmatrix}.
\end{equation}
Because of the relations of
\begin{equation}
\frac{1}{m_{\text{eff}}} =    \frac{|C|^2}{m_c}+ \frac{|X|^2}{m_X} \sim\frac{|C|^2}{m_c}, 
\end{equation}
and
\begin{equation}
V_{\text{eff}} = |C|^2 V_0,
\end{equation}

the complete Hamiltonian $\bm{H}$ is written as
\begin{eqnarray}
	\bm{H} & = & \bm{\varepsilon}_X + \begin{bmatrix} 
		\bm{\Delta} & \bm{g_0} \\
		\bm{g_0} & \bm{0}\\
	\end{bmatrix}
	+ \frac{1}{|C|^2}\begin{bmatrix} 
		\bm{H}_1 - \bm{\varepsilon} & \bm{0} \nonumber\\
		\bm{0} & \bm{0}\\
	\end{bmatrix}\\
	& = &  \bm{\varepsilon}_X  - \frac{\bm{\varepsilon}}{|C|^2} + \Delta\begin{bmatrix} 
		\bm{1} & \bm{\frac{g_0}{\Delta}} \\
		\bm{\frac{g_0}{\Delta}} & \bm{0}\\
	\end{bmatrix} + \frac{1}{|C|^2}\tilde{\bm{H}}_1.
\end{eqnarray}

In the cavity-photon limit ($\Delta \rightarrow - \infty$), which is the extreme red-detuned case, we have $|C|^2 \rightarrow 1$  and $\frac{g_0}{\Delta} \rightarrow 0^{-}$, yielding that the complete Hamiltonian can be further approximated to 
\begin{equation}
\bm{H} =
\begin{bmatrix} 
\bm{\varepsilon}_X -\bm{ \varepsilon + \Delta }+ \bm{H}_1 & \bm{0}\\
\bm{0}& \bm{0}\\
\end{bmatrix}.
\end{equation}
Therefore,  if we set $\varepsilon = \varepsilon_X + \Delta$, the complete Hamiltonian $\bm{H}$ is just the approximated  $\tilde{\bm{H}}$ 

\renewcommand{\theequation}{B-\arabic{equation}}
\renewcommand{\thefigure}{B\arabic{figure}}
\renewcommand\figurename{Figure}

\section{Tight-Binding model theory and the fitting for  $t$ and $t^\prime$}\label{app:tb}

The standard tight-binding (TB) model is adopted from the energy band structure calculation of graphene to extract the two dominant hopping integrals of Bloch exciton-polaritons in the honeycomb lattice potentials.~\cite{kasprzak2006bose} We consider the nearest-neighbor hopping integral $ t $ and next-nearest-neighbor hopping integral in the honeycomb. The corresponding tight-binding Hamiltonian is,
\begin{widetext}
\begin{equation}
 H=\sum\limits_{j}^{}\left( t \sum\limits_{|i-j|=d}^{} c_{Bi}^\dagger c_{Aj}+h.c.+t'\sum\limits_{|i-j|=\sqrt{3}d}^{}c_{Ai}^\dagger c_{Aj}+t'\sum\limits_{|i-j|=\sqrt{3}d}^{}c_{Bi}^\dagger c_{Bj}+E_0 c_j^\dagger c_j \right),
\end{equation}
\end{widetext}
where the $ i $ in the first term denotes the nearest-neighbor site to $ j $ and $ i $ in the third and fourth term denote the next-nearest-neighbor site to $ j $, $ E_0 $ indicates the on-site energy. The Fourier transformation of $ c_{Aj} $ is
\begin{equation}\label{key}
c_{Aj}=\int\limits_{\Omega}^{}e^{-i \bm{k}\cdot\bm{r_j}}c_{A\bm{k}}\frac{d \bm{k}}{(2\pi)^2}, 
\end{equation}
where $ \omega $ is the area of Brillouin zone. Similarly, we can write down the same expression for $ c_{Bi} $ and substitute them into the tight-binding Hamiltonian:
\begin{widetext}
\begin{equation}
 H=\sum\limits_{j}^{}t \int \sum\limits_{|i-j|=d}^{}e^{i \bm{k'}\cdot(\bm{r_j}+\overrightarrow{AB}_i)-i\bm{k}\cdot\bm{r_j}}c_{B\bm{k'}}^\dagger c_{A\bm{k}} +h.c.+\sum\limits_{|i-j|=\sqrt{3}d}^{} (E_0+t' e^{i \bm{k'}\cdot(\bm{r_j}+\overrightarrow{AA}_i)-\bm{k}\cdot\bm{r_j}})(c_{A\bm{k'}}^\dagger c_{A\bm{k}} +c_{B\bm{k'}}^\dagger c_{B\bm{k}} )\frac{d \bm{k}}{(2\pi)^2}\frac{d \bm{k'}}{(2\pi)^2} 
\end{equation}
\end{widetext}
Note that $ \sum\limits_{j}^{}e^{i(\bm{k'}-\bm{k})\cdot\bm{r_j}}=(2\pi)^2\delta(\bm{k}-\bm{k'}) $, therefore, the tight-binding Hamiltonian is simplified to
\begin{widetext}
\begin{equation}
H=\begin{pmatrix}
c_{A\bm{k}}^\dagger & c_{B\bm{k}}^\dagger
\end{pmatrix}\begin{pmatrix}
E_0+t' \sum\limits_{|i-j|=\sqrt{3}d}^{} e^{i \bm{k}\cdot\overrightarrow{AA}_i} & t \sum\limits_{|i-j|=d}^{} e^{i \bm{k}\cdot \overrightarrow{AB}_i}\\
t \sum\limits_{|i-j|=d}^{} e^{-i \bm{k}\cdot \overrightarrow{AB}_i} & E_0+t' \sum\limits_{|i-j|=\sqrt{3}d}^{} e^{i \bm{k}\cdot\overrightarrow{BB}_i}
\end{pmatrix}
\begin{pmatrix}
c_{A\bm{k}}\\c_{B\bm{k}}
\end{pmatrix}.
\end{equation}
\end{widetext}
Substituting the nearest-neighbor vector $ \overrightarrow{AB}_i=\left(0, d\right)$, $\left(\frac{\sqrt{3}}{2}d,\frac{d}{2}\right)$, $\left(-\frac{\sqrt{3}}{2}d,\frac{d}{2}\right)$ and  next-nearest-neighbor vector $ \overrightarrow{AA}_i=\overrightarrow{BB}_i=\left(\sqrt{3}d,0\right)$, $\left(-\sqrt{3}d,0\right)$, $\left(\frac{\sqrt{3}}{2}d,\frac{3}{2}d\right)$, $\left(-\frac{\sqrt{3}}{2}d,\frac{3}{2}d\right)$, $\left(\frac{\sqrt{3}}{2}d,-\frac{3}{2}d\right)$, $\left(-\frac{\sqrt{3}}{2}d,-\frac{3}{2}d\right) $ and summing them up, we obtain the energies(without loss of generality, we assume $ t $ positive)
\begin{equation}\label{key}
 E = E_0 \pm t\sqrt{3+f(\bm{k})}- t^\prime f(\bm{k}),
\end{equation}
where
\begin{equation}
f(\bm{k}) = 2 \cos(\sqrt{3}k_xd)+4\cos\Big(\frac{\sqrt{3}}{2}k_xd \Big)\cos\Big(\frac{3}{2}k_yd \Big).
\end{equation}
In experimental data analysis, we first obtain the energy values at all $k_y$ from the intensity peaks by aforementioned Lorentzian fitting. For the $d$ = 3 $\mu$m devices, we use a model with two Lorentzian shapes overlapped to find the peaks from the experimental linewidth plots, which inscribes the energy of two states in the $s$-band. For the $d$ = 4 $\mu$m devices, we only use one Lorentzian shape to find the $ s $-band because the lower state is mostly populated in the first BZ, whilst the higher state is populated in the second BZ such that can barely be seen in the first BZ. 

After the Lorentzian fitting to find the peak, the TB model is applied to fit the four parameters: the offset energy $E_0$, site-to-site distance $d$, nearest-neighbor hopping integral $t$ and next-nearest-neighbor hopping integral $t^\prime$. In addition to the best fitting parameters, we also present (1) standard deviation, which is the average deviation between the experimental point and that fitted point among all momentum $k_y$; (2) P-value, which can be understood to nullify the hypothesis if P-value $\gtrsim 0.05$, otherwise we can reject the null hypothesis if P-value $\ll 0.05$. We want the small P-value to accept hypothesis such that the fitting model is convincing; (3) error bar, which is set to 95$\%$ confidence intervals. The boundary of confidence intervals delimits the error range. 

\begin{figure*}[t]
	\centering  \includegraphics[width=14cm]{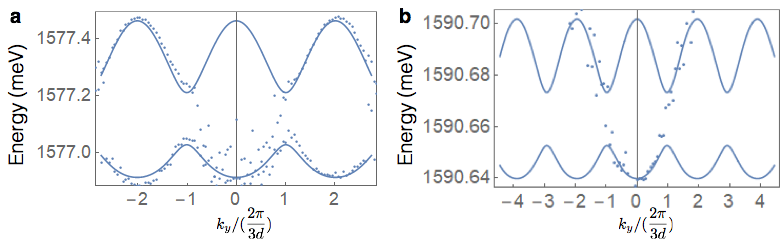}
	\caption{The fitting model (solid line) and experimental peaks (dots) of a red-detuned device  $d$ = 3 $\mu$m, $\Delta = -14.6$ meV (a) and a blue-detuned device of $d$ = 4 $\mu$m, $\Delta =  8.7$ meV (b). The $x$-axis is momentum in the unit of ${\mu m}^{-1}$ and the $y$-axis is energy in the unit of meV. 
	}
	\label{fig:B1}
\end{figure*}

We present the examples of the fitting results for the $d$ = 3 $\mu$m and $d$ = 4 $\mu$m devices in Figs.~\ref{fig:B1} (a) and (b). The fitting parameters for specific data from the $d$ = 3 $\mu$m and $d$ = 4 $\mu$m devices are collected in Table 1 and 2 respectively. Two lowest bands of Eq. (B-5) along the line 1, $\it{\Gamma}-M$ specified in Fig.~\ref{fig:1}(c) are drawn in straight line, where the dots are taken from the maximum intensity peaks in the measured experimental bandstructures. Due to our spectral linewidth, we are not able to identify small gaps between the first and the second bands along the line 1, $\it{\Gamma}-M$ unlike the theoretical plots; however, the curvature of the bands directly quantifies the hopping integrals $t$ and $t^\prime$. Note that the gap between two bands closes at $K$ and $K^\prime$ points, forming the famous linear gapless Dirac dispersions, while the curvature of two bands remains as same. Therefore, our TB dispersion fitting to the experimental data along the $\it{\Gamma}-M$ line is still valid to extract the values of $t$ and $t^\prime$.

\begin{table*}[htbp]
	\centering
		\begin{tabular}{cccccc}
			\hline
			\hline
			& \text{Best Fit} & \text{Standard deviation} & \text{P-value} & \text{Error bar(min)} & \text{ Error bar(max)}\\
			\hline
			\hline
			$E_0$/meV & 1577.14 & 0.003528 & 2.43E-923 & 1577.14 & 1577.15 \\
			$t/\mu$eV & -92.57 & 1.321 & 3.39E-145 & -95.18 & -89.97 \\
			$t^\prime/\mu$eV & 7.84319 & 1.01 & 3.89E-13 & -9.84 & 5.85 \\
			$d/\mu$m & 2.88884 & 0.020997& 6.58E-204& 2.84744&2.93024\\
			\hline
			\hline
		\end{tabular}
		\caption{A red-detuned device of $d$ = 3 $\mu$m, $\Delta = -14.6$ meV.}
\end{table*}
	
\begin{table*}[t]
	\centering
	\begin{tabular}{cccccc}
			\hline
			\hline
			& \text{Best Fit} & \text{Standard deviation} & \text{P-value} & \text{Error bar(min)} & \text{ Error bar(max)}\\
			\hline
			\hline
			$E_0$/meV & 1582 & 0.000507 & 1.48E-556 & 1582 & 1582\\
			$t/\mu$eV & -9.41 & 0.197 & 3.45E-71 & -9.8 & -9.02\\
			$t^\prime/\mu$eV & 0.730674 & 0.141 & 1.18E-6 & 0.45 & 1.011 \\
			$d/\mu$m & 3.72502 & 0.043485& 4.056E-96 & 3.63876 & 3.81128 \\
			\hline
			\hline
	\end{tabular}
	\caption{A blue-detuned device of $d$ = 4 $\mu$m, $\Delta = 8.7$ meV.}
\end{table*}
	
\bibliography{QuS.bib}

\end{document}